\documentclass[12pt,a4paper]{article}
\usepackage{amssymb}
\begin{document}
\global\parskip 6pt


\begin{titlepage}
\hfill{DIAS-STP-04.17}

\hfill{hep-th/0411185} \vspace*{1cm}
\begin{center}
{\Large\bf Pre-Big Bang Scenario on Self-T-Dual Bouncing Branes}\\

\vspace*{2cm} Massimiliano Rinaldi\footnote{E-mail: {\tt
massimiliano.rinaldi@ucd.ie}} and
Paul Watts\footnote{E-mail: {\tt watts@stp.dias.ie}}\\

\vspace*{.5cm} {\it ${}^{1,2}$Department of Mathematical Physics,
University College Dublin,\\
Belfield, Dublin 4,
Ireland}\\
\vspace{.1in} {\it ${}^2$School of Theoretical Physics,
Dublin Institute for Advanced Studies,\\
10 Burlington Road, Dublin 4,
Ireland}\\
\vspace{2cm}

\begin{abstract}
\noindent We consider a new class of 5-dimensional dilatonic actions
which are invariant under T-duality transformations along three
compact coordinates, provided that an appropriate potential is
chosen. We show that the invariance remains when we add a boundary
term corresponding to a moving 3-brane, and we study the effects of
the T-duality symmetry on the brane cosmological equations.  We find
that T-duality transformations in the bulk induce scale factor
duality on the brane, together with a change of sign of the pressure
of the brane cosmological matter. However, in a remarkable analogy
with the Pre-Big Bang scenario, the cosmological equations are
unchanged. Finally, we propose a model where the dual phases are
connected through a scattering of the brane induced by an effective
potential. We show how this model can realise a smooth, non-singular
transition between a pre-Big Bang superinflationary Universe and a
post-Big Bang accelerating Universe.
\end{abstract}
\end{center}
\end{titlepage}

\vspace{1.5 cm}

\section{Introduction}\setcounter{equation}{0}
Two of the main lines of investigation in string cosmology are the
Pre-Big Bang scenario and brane cosmology.  The former is
essentially based on a large symmetry group of the effective
bosonic string action which has been studied for many years,
leading to very important theoretical and phenomenological results
(for a review, see \cite{Gasperini}).  Brane cosmology is a more
recent idea which springs from the pioneering works of
Ho\u{r}ava-Witten \cite{Horava} and Randall-Sundrum
\cite{Randall1}-\cite{Randall2}. In the simplest models, our
Universe is seen as a warped brane embedded in 5-dimensional
space-time, where matter and all interactions except gravity are
confined (for reviews, see, for example,
\cite{Langlois}-\cite{Brax2}).

In this paper, we aim to explore possible connections between these
two ideas.  A first investigation was carried on by one of the
authors, in the context of type IIA and type IIB supergravity
\cite{max}.  These theories are considered as different limits of a
unique underlying theory, and solutions of one can be mapped into
solutions of the other by T-duality transformations of the fields
(see, for example, \cite{sen}).  This symmetry also holds when the
actions are compactified to 5 dimensions, at least in the case
considered in \cite{max}.  The dual 5-dimensional backgrounds were
chosen in order to study a moving brane with a homogeneous and
isotropic induced metric.  It was found that T-duality
transformations of the backgrounds induce the inversion of the brane
scale factor.  However, provided that the brane Lagrangean is
assumed to have a certain form, the cosmological equations are
unchanged, which suggests a possible analogy with the Pre-Big Bang
scenario.  Indeed, in the simplest string cosmological models, the
action is invariant under T-duality transformations.  Usually, these
take the form of an inversion of the scale factor (``scale factor
duality'') together with a change of sign of the pressure of the
cosmic matter, which leave the equations of motion unchanged
\cite{Gasperini}.  This analogy, however, is only valid at the level
of the equations of motion.  Indeed, in the Pre-Big Bang scenario,
the T-duality symmetry group also leaves the action unchanged.
However, in the case studied in \cite{max}, while the brane
equations of motion are invariant under T-duality, the action
transforms from type IIA to type IIB (or vice versa).

In light of these results, it is natural to ask what happens when
the brane moves in a background which is a solution of an action
invariant under T-duality.  In this paper, we address this
question by first finding such backgrounds.  This is not as easy
as it might appear, because, together with a self-T dual action,
we also need a background such that the induced metric on the
brane is homogeneous and isotropic.  In Sec.~2, we present a
dilatonic action and a metric which meet these requirements. These
solutions appear to be new and interesting in their own right, not
only in the context of brane cosmology. In this Section we
consider one simple solution to the bulk equations of motion, but
we believe that more general ones can be found.

In Sec.~3, we study the Israel junction conditions which arise when
the 3-brane is embedded in the bulk.  We first show that the
addition of the brane action does not spoil the T-duality invariance
of the total action.  Then, we obtain the brane cosmological
equations and we show that T-duality transformations in the bulk
induce the inversion of the scale factor on the brane together with
a change of sign of the pressure.  By assuming a specific form for
the brane-matter Lagrangean, we also show that by choosing an
appropriate conformal frame on the brane, the energy is conserved.

In Sec.~4, we consider the cosmological equations in the case when
the bulk background is described by the solutions found in Sec.~2.
Even in this simple case, we will see that it is not possible to
find an exact solution, mainly because the equation of state
relating energy density and pressure is manifestly dependent on the
position of the brane.  However, we will be able to show that, at
least in the regimes of late and early times, we recover the main
features of most brane cosmological models.

In Sec.~5, we consider a case where the brane scatters against the
zero of an effective potential.  This model not only strengthens the
similarities to the Pre-Big Bang scenario, but also offers a natural
interpretation of the dual solutions to the cosmological equations.
By assuming that the dual transition takes place at the bounce, we
show that there is a transition from a superinflationary phase to a
post-big bang phase, which appears to be smooth and non-singular.
Finally, we conclude with some remarks and open problems.

\section{Self-T-Dual Backgrounds}\setcounter{equation}{0}
\noindent In this Section we introduce a new family of tensor-scalar
actions which show a non-trivial invariance under field
transformations.  To begin with, consider a 5-dimensional
pseudo-Riemannian manifold ${\cal M}$ (the bulk space-time),
equipped with a metric tensor $g_{AB}$ whose line element reads
\begin{equation}\label{metric}
ds^2=g_{AB}\,dx^Adx^B=-A^2(r)\,dt^2+B^2(r)\,dr^2+R^2(r)\,
\delta_{ij}\,dx^i\,dx^j.
\end{equation}
We assume that the functions $A$, $B$ and $R$ of the radial
coordinate $r$ satisfy the equations of motion derived from the bulk
tensor-scalar action
\begin{equation}\label{action}
S_{\rm bulk}=\int_{\cal M}d^5x\sqrt{g}\,e^{-2\phi}\left[{\cal
R}+4(\nabla \phi)^2+V\right],
\end{equation}
where ${\cal R}$ is the Ricci scalar, $\phi$ is the dilaton field,
and the potential $V$ is some function of $\phi$ and possibly
$g_{AB}$.  Explicit solutions to the equations of motion are
well-known in the case when the dilaton is a function of $r$ only
and the potential has the Liouville form $V(\phi)=V_0\,e^{\,k\phi}$
\cite{Charm}.

We shall now show that another class of solutions exists provided
the potential is a function of the so-called shifted
dilaton\footnote{In the context of string cosmology, this kind of
potential is often called ``non-local'' \cite{Gasperini}.}, defined
as \begin{equation}\label{shiftdil}
\bar\phi(r)=\phi(r)-\frac{3}{2}\ln R(r).
\end{equation} This definition holds
provided that the volume of the spatial sections is constant at each
fixed $r$ \cite{Gasperini}. With such a potential, and a line
element of the form (\ref{metric}), the action has a non-trivial
symmetry which can be exploited to generate a new and inequivalent
solution to the equations of motion from a known one.  To show this,
we compute ${\cal R}$ in terms of the fields $A,B,R$ and $\phi$, and
write the action as
\begin{eqnarray}
S_{\rm bulk}=-2\int_{\cal
M}d^5x\frac{AR^3e^{-2\phi}}{B}\left[\,3{\cal
H}\left(\frac{A^{\,\prime}}{A}-\frac{B^{\,\prime}}{B}\right)-
\frac{A^{\,\prime}B^{\,\prime}}{AB}+6{\cal
H}^2\right.\\\nonumber\\\nonumber
\left.+\frac{A^{\,\prime\,\prime}}{A}+3{\cal
H}^{\,\prime}-2(\phi^{\,\prime})^2 -\frac{B^2V}{2}\right],
\end{eqnarray}
where the prime stands for differentiation with respect to $r$, and
we have defined ${\cal H}:=R^{\,\prime}/R$. By changing from $\phi$
to $\bar\phi$ and by using the identities
\begin{eqnarray}\nonumber
\frac{Ae^{-2\bar\phi}}{B}\left[ {\cal
H}\left(\frac{A^{\,\prime}}{A}-\frac{B^{\,\prime}}{B}\right)+{\cal
H}^{\,\prime}-2\phi^{\,\prime}{\cal H}+3{\cal
H}^2\right]&=&\frac{d}{dr}\left(\frac{A{\cal
H}e^{-2\bar\phi}}{B}\right),\\\label{identity}\\\nonumber
\frac{A'e^{-2\bar\phi}}{B}\left(\frac{A^{\,\prime\,\prime}}{A^{\,\prime}}
-\frac{B^{\,\prime}}{B}-2\bar\phi^{\,\prime}\right)&=&
\frac{d}{dr}\left(\frac{e^{-2\bar\phi}A^{\,\prime}}{B}\right),
\end{eqnarray}
we can use Stokes' theorem to write the action in the form
\begin{eqnarray}\label{redaction}
S_{\rm bulk}&=&\int_{\cal M} d^5x\,
e^{-2\bar\phi}\left[ABV(\bar\phi)-\frac{3A{\cal
H}^2}{B}+\frac{4(\bar\phi^{\,\prime})^2A}{B}-
\frac{4\bar\phi^{\,\prime}A^{\,\prime}}{B}\right]\nonumber\\
&&-2\int_{\partial\cal M}d^3x\,dt\,e^{-2\bar\phi}\left(\frac{3A{\cal
H}}{B}+\frac{A^{\,\prime}}{B}\right).
\end{eqnarray}
Note that we have also assumed that the potential $V$, which in
principle depends on both $\phi$ and $g_{AB}$, becomes a function of
{\em only} the shifted dilaton $\bar\phi$ under the metric ansatz
(\ref{metric}). Since the fields $A,B$, and $R$ are independent of
all coordinates except $r$, the metric (\ref{metric}) and the
equations of motion obtained from the action (\ref{action}) are
clearly invariant under translations along the compact coordinates
$x^i$, $i=1,2,3$. In addition, if we assume that the fields vanish
at the boundary $\partial{\cal M}$, the action (\ref{redaction}) is
invariant under the field transformations
\begin{eqnarray}\label{tdualtransf}
R(r)\longrightarrow\tilde R(r)&=&R(r)^{-1}\quad\Rightarrow\quad
{\cal H}\longrightarrow -{\cal H}\\\nonumber\\\label{tdualtransf2}
\phi(r) \longrightarrow\tilde \phi(r)&=&\phi(r)-3\ln
R(r)\quad\Rightarrow\quad \bar\phi(r)\longrightarrow \bar\phi(r).
\end{eqnarray}
Therefore, for every solution to the equations of motion with line
element (\ref{metric}) and dilaton $\phi(r)$, there exists a dual
counterpart with
\begin{eqnarray}\label{dualmetric}
d\tilde s^2=-A^2(r)dt^2+B^2(r)dr^2+R^{-2}(r)\delta_{ij}\,dx^idx^j,
\end{eqnarray}
and dilaton $\tilde\phi$.  In the context of string theory, this is
the well known T-duality symmetry and the transformations
(\ref{tdualtransf}) and (\ref{tdualtransf2}) are a particular case
of Buscher's transformations \cite{Buscher1,Buscher2} applied along
each $x^i$.

The equations of motion can obtained by variation of
(\ref{redaction}) with respect to the fields $A$, $B$, $R$ and
$\bar\phi$. If we assume that these vanish at the boundary
$\partial{\cal M}$, we find
\begin{eqnarray}\label{first}
\bar\phi^{\,\prime\,\prime}-\bar\phi^{\,\prime}\left(\frac{A^{\,\prime}}{A}+
\frac{B^{\,\prime}}{B}\right)-\frac{3}{2}{\cal H}^{\,2}&=&0\\
\nonumber\\
\label{second}\bar\phi^{\,\prime\,\prime}-\bar\phi^{\,\prime}\frac{{\cal
    H}^{\,\prime}}{{\cal H}}+\frac{1}{2}B^2V(\bar\phi)&=&0\\\nonumber\\
\label{third}\frac{d}{dr}\left[\ln\left(\frac{A{\cal
      H}e^{-2\bar\phi}}{B}\right)\right]&=&0\\\nonumber\\
\frac{\partial
  V(\bar\phi)}{\partial \bar\phi}-4V(\bar\phi)-\frac{4{\cal
    H}}{B^2}\frac{d}{dr}\left[\frac{1}{{\cal H}}\left(\frac{{\cal
      H}^{\,\prime}}{{\cal
      H}}-\frac{B^{\,\prime}}{B}\right)\right]&=&0.
\end{eqnarray}
However, it can be shown that if $\bar\phi^{\,\prime}\neq 0$, the
last of these equations is just a combination of the other three.
Therefore we are left with three equations for the five unknown
functions $A$, $B$, $R$, $V$ and $\bar\phi$.  If one defines new
functions $N(r)$ and $P(r)$ by
\begin{equation}
N:=\frac{Ae^{-2\bar\phi}}{B},\quad P:=\frac{1}{AB},
\end{equation}
then (\ref{first})-(\ref{third}) can be rewritten as
\begin{eqnarray}\label{motionMN1}
\bar\phi^{\,\prime\,\prime}+\frac{P'}{P}\bar\phi^{\,\prime}-\frac{3}{2}{\cal
H}^{\,2}&=&0,\\
\nonumber\\
\label{motionMN2} N{\cal H}&=&k,\\\nonumber\\\label{motionMN3}
\bar\phi^{\,\prime\,\prime}-\frac{{\cal H}^{\,\prime}}{{\cal
H}}\bar\phi^{\,\prime}+\frac{V(\bar \phi)}{2NP}\,e^{-2\bar\phi}&=&0,
\end{eqnarray}
where $k$ is an arbitrary constant.

We found a large number of singular and non-singular solutions to
these equations, which often display an unusual asymptotic
structure, i.e.\ they are neither asymptotically flat nor de Sitter
or anti-de Sitter.  These will be discussed in detail elsewhere
\cite{upcoming}.  For the moment, we consider a simple solution
which will be useful later: if we assume that $R(r)=r$ and that the
shifted dilaton has the form
\begin{equation}\label{dil}
\bar\phi=\alpha\ln r-\beta\ln P,
\end{equation} where $\alpha$ and $\beta$ are constants, then Eq.~(\ref{motionMN1}) yields
\begin{equation}
P^{\,\prime\,\prime}-\frac{\alpha}{\beta
r}P^{\,\prime}+\frac{(2\alpha +3)}{2\beta r^2}P=0.
\end{equation}
The general solution to this is
\begin{equation}\label{pi}
P(r)=p_1r^{m_1}+p_2r^{m_2},
\end{equation}
where $p_1$ and $p_2$ are integration constants, and
\begin{equation}\label{emme}
m_{1,2}=\frac{1}{2\beta}\left[\alpha+\beta\pm\sqrt{(\alpha-\beta)^2
-6\beta}\right].
\end{equation}
Then, it follows that
\begin{eqnarray}\label{gtt}
A^2&=&k\,r^{\,2\alpha+1}P^{\,-2\beta-1},\\\nonumber\\\label{grr}
B^2&=&k^{-1}r^{\,-2\alpha-1}P^{\,2\beta-1},\\\nonumber\\\label{potential}
V(\bar\phi)&=&\frac{2kp_1p_2}{\beta}\left[(\alpha-\beta)^2-6\beta\right]
e^{\frac{(2\beta-1)} {\beta}\,\bar\phi},\\\nonumber\\
\phi&=&\left(\alpha+\frac{3}{2}\right)\ln r-\beta\ln P.
\end{eqnarray}
The T-duality symmetry of the action is manifest in the equations of
motion (\ref{motionMN1})-(\ref{motionMN3}), which are indeed
invariant under the transformations\footnote{Note that, under the
transformation (\ref{tdualtransf}), we also have that $k\rightarrow
-k$ in Eq.~(\ref{motionMN2}).  However, given that $k$ is arbitrary,
we shall ignore this detail.} (\ref{tdualtransf}) and
(\ref{tdualtransf2}). Hence, to the solution above it corresponds a
new and inequivalent one obtained by replacing $R=r$ with $\tilde
R=1/r$ and $\phi$ with $\phi-3\ln r$.

Let us analyse this solution and its dual counterpart more
closely: The Ricci scalar has the form
\begin{equation}\label{ricciscal}
{\cal
R}=-\frac{2}{r^2B^2}\,\left[c_2\left(\frac{rP^{\,\prime}}{P}\right)^2
+c_1\left(\frac{rP^{\,\prime}}{P}\right)+c_0\right],\end{equation}
where $c_2$, $c_1$, and $c_0$ are functions of $\alpha$ and $\beta$.
The T-dual Ricci scalar differs only in the value of the constant
$c_0$.  Given the form of $P$ and $B^2$ from (\ref{pi}) and
(\ref{grr}) respectively, it is clear that both Ricci scalars are
finite at infinity provided that the term $r^2B^2$ tends to a
constant for large $r$.  By imposing this constraint, we find an
implicit relation between $\alpha$ and $\beta$, namely
\begin{equation} m_1=\frac{2\alpha-1}{2\beta-1},\end{equation} where
we have taken $m_1$ to be the largest of the two solutions
(\ref{emme}).  Furthermore, provided $\beta<-1/2$, a regular event
horizon exists and the Ricci scalar diverges only at $r=0$.  The
$g_{tt}$ and $g_{rr}$ components of the metric (\ref{metric}) and of
its T-dual (\ref{dualmetric}) can be written as
\begin{eqnarray}
g_{tt}&=&k\,r^{\,2(1-m_1)}\,p_1^{\,-(1+2\beta)}\left[1-\left(\frac{r_h}{r}\right)^{\gamma}\,
\right]^{\,-(1+2\beta)},\\\nonumber\\
g_{rr}&=&k^{-1}r^{-2}p_1^{\,(2\beta-1)}\left[1-\left(\frac{r_h}{r}\right)^{\gamma}\,
\right]^{\,(2\beta-1)},
\end{eqnarray}
where we have defined $\gamma=m_1-m_2$, and
\begin{equation}
r_h=\left(-\frac{p_2}{p_1}\right)^{1/\gamma}
\end{equation}
determines the location of the horizon, which is clearly invariant
under T-duality.  With these conventions, we have $\gamma>0$,
$k>0$, $p_1>0$, and $p_2<0$.  At large $r$, the metrics
(\ref{metric}) and (\ref{dualmetric}) become, respectively,
\begin{eqnarray} ds^2&\simeq&
-r^{2(1-m_1)}\,dt^2+r^{-2}\,dr^2+r^2\delta_{ij}\,dx^idx^j,\\\nonumber\\
d\tilde{s}^2&\simeq&
-r^{2(1-m_1)}\,dt^2+r^{-2}(dr^2+\delta_{ij}\,dx^idx^j),\end{eqnarray}
(where we have ignored multiplicative constants).  The analysis of
the Ricci and Riemann tensors reveals that these space-times are
neither flat, nor Ricci flat, nor (anti)-de Sitter.  In
particular, note that the T-dual metric is the product between the
time line and a 4-dimensional hyperbolic space.  We also note
that, at large $r$, the shifted dilaton diverges, with the same
sign as the expression $\alpha-\beta m_1$.  Accordingly, the
potential $V(\bar\phi)$, which is positive everywhere, vanishes at
infinity when the latter is negative, and diverges otherwise.
 Finally, both dilaton field and potential vanish at the horizon
for any value of $\alpha$ and for all $\beta<-1/2$.

\section{T-Duality and Junction Conditions}\setcounter{equation}{0}
\noindent We now investigate the properties of a $3$-brane $\Sigma$
embedded in the bulk space-time ${\cal M}$.  In particular, we focus
on a moving 3-brane with an induced metric of the form
\begin{equation}\label{indmetric}
ds^2_{\Sigma}=-d\tau^2+R^2(r)\,\delta_{ij}\,dx^i\,dx^j,
\end{equation}
where the cosmic time $\tau$ will be defined shortly.  Similar
models have been widely studied, and it is well-known that an
observer living on the moving brane will experience an evolving
4-dimensional Universe \cite{Langlois}-\cite{Brax2}. In particular,
we are interested on the effects that the T-duality symmetry of the
bulk might have on the brane cosmological equations.

The presence of the 3-brane introduces the extra term in the action
\begin{equation}\label{extraterm}
S_{\rm
brane}=-\int_{\Sigma}d^3x\,d\tau\sqrt{h}e^{-2\phi}\left[2\left(K^++K^-\right)+{\cal
L}\right],
\end{equation}
where $K^{\pm}$ are the extrinsic curvatures on the two sides of the
brane.  ${\cal L}$ represents the Lagrangean of the matter confined
on the brane and $h$ is the determinant of the induced metric. If
this has the form (\ref{indmetric}), then $\sqrt{h}=R^3$ and we have
the relation $\sqrt{h}\,e^{-2\phi}=e^{-2\bar\phi}$.

If we assume a ${\mathbb Z}_2$ symmetry around the brane then
$K^+=K^-=K$, and the junction conditions (in the string frame
\cite{max,Foffa}) read
\begin{eqnarray}\label{junction1}
2K_{\mu\nu}&=&T_{\mu\nu}-\frac{1}{2}\,T^{\phi}h_{\mu\nu},\\\nonumber\\
\label{junction2} 4n^A \nabla_A\phi&=&T-\frac{3}{2}\,T^{\phi}.
\end{eqnarray}
In these equations, $T_{\mu\nu}$ defines the energy-momentum tensor
of the brane matter
\begin{equation}\label{oldemtensor}
T_{\mu\nu}=-\frac{1}{\sqrt{h}}\frac{\delta(\sqrt{h}{\cal L})}{\delta
  h^{\mu\nu}},
\end{equation}
$T^{\phi}$ is the variation of the Lagrangean with respect to the
scalar field
\begin{equation}
T^{\phi}=-\frac{1}{2}e^{2\phi}\,\frac{\delta (e^{-2\phi}{\cal
L})}{\delta \phi},
\end{equation}
and $n^A$ is the unit vector normal to the brane pointing into the
bulk\footnote{Here we adapt the conventions of \cite{Foffa}, where
the dilaton coupling in the action is $\exp(-\phi)$ instead of
$\exp(-2\phi)$.}. If the bulk and induced metrics are (\ref{metric})
and (\ref{indmetric}) respectively, then the conformal time $\tau$
is implicitly defined by the normalisation condition
\begin{equation}\label{normalization}
A\dot t=\sqrt{1+B^2\dot r^2},
\end{equation} where the dot stands for differentiation with respect
to $\tau$. Choosing the positive, rather than the negative, square
root here ensures that $\tau$ increases with the bulk coordinate
time $t$, i.e.\ that the unit normal vector points {\em into} the
bulk \cite{Chamblin}.  With these conventions, the components of the
normal vector and of the extrinsic curvature read \cite{Brax}
\begin{eqnarray}\label{normal}
n_t&=&AB\dot r,\quad n_r=-B\sqrt{1+B^2\dot r^2},\\\nonumber\\
\label{extrinsic} K_{ij}&=&-\frac{R^{\,\prime}}{BR}\sqrt{1+B^2\dot
r^2}\,h_{ij}, \quad
K_{\tau\tau}=\frac{1}{AB}\frac{d}{dr}\left(A\sqrt{1+B^2\dot
r^2}\right).
\end{eqnarray}
Normally, the conditions (\ref{junction1}) and (\ref{junction2})
give the Friedmann equation on the brane, together with an energy
(non-)conservation equation and a junction condition for the dilaton
\cite{Langlois}-\cite{Brax2}.

In light of the T-duality invariance of the bulk action, it is
interesting to investigate how the junction conditions behave under
T-duality transformations in the bulk. Indeed, it is worth noting
that the transformation (\ref{tdualtransf}) implies
\begin{equation}
h_{ij}\stackrel{T}{\longrightarrow}\tilde
h_{ij}=\frac{1}{R^4(r)}h_{ij}=\frac{1}{R^2(r)}\delta_{\,ij}\,,
\end{equation}
which means that the scale factor of the induced metric
(\ref{indmetric}) undergoes inversion under T-duality.  A very
similar situation occurs in the context of string cosmology, where
T-duality induces the inversion of the scale factor (scale factor
duality) and leaves the cosmological equations unchanged, provided
that $p\rightarrow -p$, where $p$ is the pressure of the matter
\cite{Gasperini}. Therefore it is natural to ask wether also in our
case the cosmological equations are left invariant by T-duality
transformations, provided some transformation law is assumed for the
brane matter. In fact, it is not hard to see that the junction
conditions (and hence the cosmological equations) are not invariant
under T-duality, unless some transformation rules are assumed for
the terms $T_{\mu\nu}$ and $T^{\phi}$.  Indeed, given that under
T-duality $\phi \rightarrow \phi-3\ln R$, the second junction
condition will contain an extra term in the derivative. Moreover,
the $ij$ components of the extrinsic curvature transform according
to
\begin{equation}
K_{ij}\stackrel{T}{\longrightarrow}\tilde
K_{ij}=-\frac{1}{R^4(r)}K_{ij},
\end{equation}
hence the right hand side of the $ij$ component of the first
junction condition changes sign.

One can argue that the T-duality invariance is broken once we add
the brane term (\ref{extraterm}) to the bulk action. Indeed, given
the components (\ref{extrinsic}) of $K_{\mu\nu}$, we see that the
trace $K$ which appears in the brane action transforms as
\begin{equation}
K\stackrel{T}{\longrightarrow}\tilde K=K-2K_{ij}h^{ij}.
\end{equation}
On the contrary, we can show that the {\em total} action is in fact
T-duality invariant.  To do so, we recall that the bulk action was
reduced to the form (\ref{redaction}), and we assumed that the
second integral, i.e.\ the boundary term, vanished on $\partial\cal
M$. However, the presence of $\Sigma$ results in an additional
boundary for the bulk space, and the boundary term will not in
general vanish at the location of the brane.  We therefore have to
add its contribution to (\ref{extraterm}). Therefore, the sum of
\emph{all} boundary terms becomes
\begin{equation}
S_{\rm boundary}=-2\int_{\partial\cal
M}d^3x\,dt\,e^{-2\bar\phi}\left[\,
\frac{3AR^{\,\prime}}{BR}+\frac{A^{\,\prime}}{B}\,\right]-\int_{\Sigma}d^3x\,
d\tau\,e^{-2\bar\phi}\left[\,2(K^++K^-)+\cal L\,\right].
\end{equation}
Recall that $\partial\cal M=\partial\cal M_+\cup\partial\cal M_-$,
where $\partial\cal M_{\pm}$ are the boundaries of the bulk on
either side of the brane.  As submanifolds of $\cal M$, these are
both equivalent to the brane $\Sigma$, but are oriented in opposite
directions (one with normal vector $n^A$, the other with $-n^A$).
The ${\mathbb Z}_2$ symmetry, however, identifies these two
boundaries with each other. Therefore, once the ${\mathbb Z}_2$
symmetry is implemented, the boundary term reads \begin{equation}
S_{\rm boundary}^{\,{\mathbb Z}_2}=-\int_{\Sigma}d^3x\,d\tau
e^{-2\bar\phi}\left[\,4\dot
t\left(\frac{3AR^{\,\prime}}{BR}+\frac{A^{\,\prime}}{B}\right)+4K+{\cal
L}\,\right],\end{equation} where we set $K^-=K^+=K$, and where we
transformed the integral over $\partial \cal M$ into an integral
over $\Sigma$ by setting
\begin{equation}
dt=\frac{dt}{d\tau}d\tau=\dot t\,d\tau.
\end{equation}
Then, by using the normalization condition (\ref{normalization}),
and the components of the extrinsic curvature, the boundary term
reduces to
\begin{equation}
S_{\rm boundary}^{\,{\mathbb
Z}_2}=-\int_{\Sigma}d^3x\,d\tau\,e^{-2\bar\phi}\left[\,{\cal L}-
\frac{4}{B}\frac{d}{dr}(A\dot t)\,\right].
\end{equation}
This term, and hence the total action $S=S_{\rm bulk}+S_{\rm
brane}$, is invariant under the transformations (\ref{tdualtransf})
and (\ref{tdualtransf2}), provided that
\begin{equation}
{\cal L}\stackrel{T}{\longrightarrow}\tilde{\cal L}={\cal L}.
\end{equation}
Despite its simple form, this transformation is far from obvious,
because ${\cal L}$ could be a complicated function of the matter
fields, the induced metric and the dilaton.  However, some insight
can be obtained by imposing the invariance of the junction
conditions and studying how the brane-matter energy-momentum tensor
behaves.

To start with, we first assume that the energy-momentum tensor for
the brane matter has the form $T^{\mu}_{\;\;\nu}={\rm
diag}(-\mu,p,p,p)$. Then we use Eq.~(\ref{junction1}) to obtain the
expression
\begin{equation}
\frac{3}{2}T^{\phi}=-2K_{ij}h^{ij}+T_{ij}h^{ij},
\end{equation}
which we insert into Eq.~(\ref{junction2}), obtaining
\begin{equation}
4n^A\nabla_A(\bar\phi+\frac{3}{2}\ln
R)=T+2K_{ij}h^{ij}-T_{ij}h^{ij}.
\end{equation}
By using Eqs.~(\ref{normal})-(\ref{extrinsic}), we find that
\begin{equation}
K_{ij}h^{ij}=3n^A\nabla_A(\ln R),
\end{equation}
and the junction condition (\ref{junction2}) reads
\begin{equation}\label{juncbarphi} 4n^A\nabla_A \bar\phi=-\mu.
\end{equation} By defining the ``shifted" pressure
\begin{equation}\bar p:=p-\frac{1}{2}T^{\phi},\end{equation} we can
write the components of the junction conditions (\ref{junction1}) as
\begin{eqnarray}\label{juncij}
2K_{ij}&=&\bar p\,h_{ij},\\\nonumber\\\label{junctautau}
2K_{\tau\tau}&=&\mu+\frac{T^{\phi}}{2},
\end{eqnarray}
Finally, by inserting  the expressions for the components of the
extrinsic curvature into these equations, by using the equation of
motion (\ref{first}), and by defining $\omega$ and the Hubble
``constant'' $H$ by
\begin{eqnarray}
\omega:=-\frac{R^{\,\prime}}{2R{\bar\phi}^{\,\prime}}\,,&&H:=\frac{\dot
R}{R}\,,
\end{eqnarray}
we find that the junction conditions reduce to the three independent
equations
\begin{eqnarray}
\bar p&=&\omega\mu,\qquad
\\\nonumber\\
H^2&=&\frac{(\omega\mu)^2}{4}-\left(\frac{R^{\,\prime}}
{RB}\right)^2,\\\nonumber\\\label{noncons} \dot\mu+3H\bar
p&=&\dot{\bar\phi}\left(2\mu+T^{\phi}\right).
\end{eqnarray}
The first is the effective equation of state for a perfect fluid
confined on the brane and the second is similar to the usual
Friedmann equation.  Finally, the third reveals that the energy on
the brane is not in general conserved, as normally happens in brane
cosmology whenever there is a bulk dilaton field
\cite{Langlois}-\cite{Brax2}.  Note that under the transformation
(\ref{tdualtransf}), the equation of state is ``reflected'', i.e.\
\begin{equation}
\omega\stackrel{T}{\longrightarrow}\tilde\omega=-\omega.
\end{equation}
This proves that the cosmological equations on the brane are
manifestly T-duality invariant, provided that
\begin{equation}\label{mattercondition}
\mu\stackrel{T}{\longrightarrow}\tilde \mu=\mu, \qquad \bar
p\stackrel{T}{\longrightarrow}\tilde{\bar p}=-\bar p\quad
\Leftrightarrow \quad \omega \stackrel{T}{\longrightarrow}
\tilde\omega=-\omega.
\end{equation}
In string cosmology, the scale factor duality is always followed by
the reflection of the equation of state, which is required by the
$O(d,d)$ invariance of the action when a matter Lagrangean
describing a perfect fluid is included \cite{Veneziano1}. Therefore,
by imposing T-duality invariance on the junction conditions, we
obtain a brane cosmological model which shares the essential
features of the Pre-Big Bang scenario.

The conditions (\ref{mattercondition}) can be clarified by choosing
a specific form for ${\cal L}$.  Following \cite{grojean}, we assume
that
\begin{equation}
{\cal L}=f^2(\phi)z(\phi)L(\psi,\nabla\psi,\gamma_{\mu\nu}),
\end{equation}
where $\psi$ represents generic fields living on the brane which
couple to the bulk dilaton only through a conformal metric
$\gamma_{\mu\nu}=f(\phi)h_{\mu\nu}$.  Hence, we can define the
energy-momentum tensor with respect to the conformal metric
$\gamma_{\mu\nu}$ as
\begin{equation}
S_{\mu\nu}=-\frac{1}{\sqrt{\gamma}}\frac{\delta(\sqrt{\gamma}\,L)}{\delta
  \gamma^{\mu\nu}},
\end{equation}
where $\sqrt{\gamma}=f^2(\phi)\sqrt{h}$.  The two energy-momentum
tensors are then related by
\begin{equation}\label{newEMT}
T_{\mu\nu}=f(\phi)z(\phi)S_{\mu\nu},
\end{equation}
and if we set $S^{\mu}_{\;\;\nu}={\rm diag}(-\rho,\pi,\pi,\pi)$, we
obtain
\begin{equation}\label{newpmu}
\mu=f^2(\phi)z(\phi)\rho,\qquad p=f^2(\phi)z(\phi)\pi.
\end{equation}
In particular, with the choice $z(\phi)=e^{2\phi}$, we find that
\begin{equation}
T^{\phi}=-\frac{1}{2}f(\phi)\frac{df(\phi)}{d\phi}\,e^{2\phi}
S^{\mu}_{\;\;\mu}.
\end{equation}
Thus, by using Eq.~(\ref{newEMT}), we obtain the relation between
$T^{\phi}$ and the trace of the energy-momentum tensor $T$
\begin{equation}\label{tphi}
T^{\phi}=-\frac{T}{2}\frac{d}{d\phi}\ln f(\phi).
\end{equation}
Incidentally, this formula can be used to show that, in the
conformal frame defined by $\gamma_{\mu\nu}$, the energy on the
brane is conserved.  In terms of $p$ and $\phi$, Eq.~(\ref{noncons})
reads
\begin{equation}
\dot \mu+3H(p+\mu)=\dot\phi (2\mu+T^{\phi}).
\end{equation}
Let the line element corresponding to the metric $\gamma_{\mu\nu}$
be
\begin{equation}
ds^2_{\gamma}=-d\xi^2+E^2(r,\phi)\delta_{ij}\,dx^idx^j=f(\phi)
(-d\tau^2+R^2(r)\delta_{ij}\,dx^idx^j).
\end{equation}
Thus, $d\xi=\sqrt{f(\phi)}d\tau$ and $E(r,\phi)=\sqrt{f(\phi)}R(r)$.
By using Eq.~(\ref{newpmu}), we find that the conservation equation
reads
\begin{equation}
\dot\rho+3\frac{\dot
E}{E}(\rho+\pi)=\rho\left(2\dot{\phi}-\frac{\dot z}{z}\right),
\end{equation}
where now the dot stands for differentiation with respect to the
conformal time $\xi$.  Therefore, when we set $z(\phi)=e^{2\phi}$,
the above equation reduces to
\begin{equation}
\dot\rho+3\frac{\dot E}{E}(\rho+\pi)=0,
\end{equation}
which shows that, in the conformal frame defined by
$\gamma_{\mu\nu}$, the energy on the brane is
conserved.\footnote{Note that, in analogy with our results, the
energy on the brane in \cite{grojean} is conserved only if
$z(\phi)=1$.}

We now come back to the junction conditions (\ref{juncbarphi}),
(\ref{juncij}) and (\ref{junctautau}).  First, we note that
Eq.~(\ref{juncbarphi}) implies that, no matter what form of ${\cal
L}$ we choose, under T-duality we must have
\begin{equation}
\mu\stackrel{T}{\longrightarrow}\tilde\mu=\mu.
\end{equation}
By using Eq.~(\ref{tphi}), we can write Eqs.~(\ref{juncij}) and
(\ref{junctautau}) as
\begin{eqnarray}
2K_{ij}&=&\left(1-\frac{3}{2}\sigma^{\,\prime}\right)ph_{ij}+
\frac{1}{2}\mu\sigma^{\,\prime}h_{ij}\,,\\\nonumber\\2K_{\tau\tau}&=&
\left(1-\frac{1}{2}\sigma^{\,\prime}\right)\mu+\frac{3}{2}\,
\sigma^{\,\prime}p\,,
\end{eqnarray}
where we set $f(\phi)=e^{-2\sigma (\phi)}$ and where
$\sigma^{\,\prime}=\frac{d\sigma}{d\phi}$.  Given that, under
T-duality,
\begin{equation}
K_{ij}\stackrel{T}{\longrightarrow}\tilde
K_{ij}=-\frac{1}{R^4(r)}K_{ij}, \qquad
h_{ij}\stackrel{T}{\longrightarrow}\tilde
h_{ij}=\frac{1}{R^4(r)}h_{ij},
\end{equation}
we see that in order to preserve T-duality invariance of the
junction conditions, $\sigma(\phi)$ and $p$ would have to transform
in a (possibly very) complicated way.  But an alternative would be
to simply require $\sigma^{\,\prime}=0$ and
$p\stackrel{T}{\longrightarrow}\tilde p=-p$.  In this case, the
matter on the brane is coupled to the bulk dilaton only through the
factor $e^{-2\phi}$ which appears in the brane action
(\ref{extraterm}).

The choice $\sigma^{\,\prime}=0$ might seem to be something of a
trivial case, but we now show that it still leads to a very
interesting cosmological model.

\section{Self T-Dual Brane Cosmology}\label{sec4}\setcounter{equation}{0}
\noindent In this Section we study the brane cosmological equations
in the case when the bulk metric and potential are given by
Eqs.~(\ref{gtt}), (\ref{grr}), and (\ref{potential}) respectively.
We also have $R(r)=r$, so the shifted dilaton reads
\begin{equation}
\bar\phi(r)=\alpha\ln r-\beta\ln P(r),
\end{equation}
where $P(r)$ is given by Eq.~(\ref{pi}).  If we also require
$\sigma'$ to vanish, the junction conditions reduce to
\begin{eqnarray}
p&=&\omega\mu,\\\nonumber\\\label{redH}
H^2&=&\left(\frac{\omega\mu}{2}\right)^2-
\frac{1}{B^2r^2},\\\nonumber\\\label{conspec}\dot\mu&=&\left(2\dot{\bar\phi}
-3H\omega\right)\mu,
\end{eqnarray}
where
\begin{eqnarray}
\label{redomega}\omega=-(2r\bar\phi^{\,\prime})^{-1}.
\end{eqnarray}
Note also that in the Friedmann equation, $H^2$ depends on $p^2$
and not on $\mu^2$, as in the usual brane cosmology.  Despite
these complications, if we assume as a background the black hole
solution described in Sec.~2, Eq.~(\ref{conspec}) can be
integrated with respect to $r$,
yielding
\begin{equation}\label{expmu}
\mu=\frac{\mu_0}{2k\,\omega\, r^2B^2},
\end{equation}
where $\mu_0$ is an arbitrary positive integration constant.

The evolution of the scale factor $r=r(\tau)$ can be better
understood by introducing the effective potential \cite{Burgess}
\begin{equation}\label{effpot}
W(r)=\frac{1}{B^2}-\left(\frac{r\omega\mu}{2}\right)^2,\end{equation}
so that the Friedmann equation can be written as
$\dot{r}^2+W(r)=0$.  This equation describes a point particle
moving in a potential $W$, whose zeroes correspond to the
classical turning points.  By using Eq.~(\ref{expmu}), we can find
an explicit form for the effective potential:
\begin{equation}
W(r)=\frac{1}{B^2}\left[1-\frac{\mu_0^2}{16\,k^2\,r^2\,B^2}\right].
\end{equation}
With this expression, it can be shown that $W(r)$ has two zeroes,
one at $r=r_h$ and the other at $r_0>r_h$,
provided
\begin{equation}\label{poscond}
\mu_0^2>16k\,q,
\end{equation} where
$q:=p_1^{\,\,2\beta-1}$.  This condition also ensures that $W(r)$
is negative (and hence the Friedmann equation is well defined) for
all $r>r_0$.  Moreover, we find that $W^{\,\prime}(r_0)<0$,
$W^{\,\prime}(r_h)=0$ and there exists a maximum between $r_h$ and
$r_0$.  At large values of $r$, $W(r)\sim -r^2$ and, finally,
$W(r)$ is negative and divergent for $r\rightarrow 0^+$, which
reveals that $r_h$ is an inflection point.

It is worth noticing that the function $\omega$, expressed in
terms of $P(r)$, reads
\begin{equation}
\omega=\frac{P}{2(r\beta P^{\,\prime}-\alpha P)}.
\end{equation}
Thus, it is easy to see that it vanishes at the horizon and it
tends to a constant at large $r$.  However, $\omega$ is an
implicit function of $\tau$, so
$|\dot\omega|=|\dot{r}\omega^{\,\prime}|\ll 1$ for $r$ close to
$r_0$ as well.  Therefore, $\omega$ is approximately constant in
the neighbourhood of the turning point.  Suppose that, in this
region, $\omega\simeq 1/3$.  Then we can write
Eq.~(\ref{redomega}) as $2\dot{\bar\phi}=-3H$, and the energy
conservation equation Eq.~(\ref{conspec}) can be immediately
integrated to give $\mu\sim r^{-4}$.  Therefore, an observer
living on the brane observes a small (compared to $r_0$)
radiation-dominated Universe.  To this solution corresponds a
T-dual counterpart with the scale factor $\tilde
R(\tau)=1/r(\tau)$ and $\tilde\omega=-\omega$.  Therefore, a
``dual'' observer sees a large Universe filled with matter with
$\tilde\omega=-1/3$, which typically corresponds to a gas of
stretched strings (see references in \cite{Gasperini}).

We now consider the Universe as seen in the present (i.e.\
matter-dominated) epoch, and take the pressure $p$ to be
vanishing. Therefore, we are left with only two independent
junction conditions:
\begin{equation}
2K_{\tau\tau}=\mu, \qquad 4n^A\nabla_A\bar\phi=-\mu,
\end{equation}
which once again yield Eqs.~(\ref{redH}) and (\ref{conspec}).  If
we assume that the present Universe is modelled by the brane
moving in the large $r$ region, then the function $\omega$ is a
constant, but it is no longer interpreted as the proportionality
factor in the equation of state.  It follows from
Eq.~(\ref{expmu}) that
\begin{equation}
\mu_{\infty}:=\lim_{r\rightarrow \infty}\mu(r)=\left(m_1\beta
-\alpha\right)\frac{\mu_0}{q}.
\end{equation}
Therefore, in the large-$r$ region we can express the energy
density as $\mu=\mu(\tau)+\mu_{\infty}$.  The constant
$\mu_{\infty}$ can be interpreted as the tension of the brane
${\cal T}$; hence, the first term on the right-hand side of
(\ref{redH}) can be expanded to first order in $\mu(\tau)$
\cite{Langlois}-\cite{Brax2}, yielding
\begin{equation}
H^2=\frac{\mu(\tau){\cal
T}}{36}+\frac{1}{q^2}\left(\frac{\mu_0^2}{16}-kq\right),
\end{equation}
where we assumed again that $\omega=1/3$ and that $r^2B^2\simeq
q/k$ at large $r$.  Given the condition (\ref{poscond}), we see
that at large $r$ we obtain a standard Friedmann equation, which
describes an expanding, accelerating Universe with positive
cosmological constant.  Like in the case of a radiation-dominated
Universe, this solution also has a dual counterpart with small
scale factor and vanishing pressure.  We now provide a possible
interpretation of these dual phases.

\section{Bouncing Branes}\setcounter{equation}{0}
\noindent In this section, we explore the possibility that the
transition between T-dual phases is triggered by the scattering of
the brane against a zero of an effective potential.  Let us consider
a general Friedmann equation written in terms of an effective
potential $W(r)$, namely
\begin{eqnarray}\label{tpoint}
\dot r^2+W(r)&=&0,
\end{eqnarray}
and assume that there exists $r_0>0$ such that $W(r_0)=0$ and
$W^{\,\prime}(r_0)<0$; this means that a brane moving in from the
region $r>r_0$ with $\dot r<0$ will bounce off the effective
potential at $r=r_0$ and move in the direction of increasing $r$,
i.e.\ $\dot r>0$.  We refer to the former as the ``pre-bounce''
epoch and the latter as the ``post-bounce'' epoch, with the bounce
itself taking place at cosmic time $\tau=0$.  The acceleration
$\ddot r=-W^{\,\prime}/2$ is always positive, at least in the region
near $r_0$.

Note that the effective potential and the position of its zeroes are
unchanged by T-duality transformations, so that the form of $W(r)$
and Eq.~(\ref{tpoint}) hold for both $R(\tau)=r(\tau)$ and the dual
$\tilde R(\tau)=1/r(\tau)$. We can utilise this property to ensure
an always-expanding universe: recall that $R$ is the scale factor
for the spatial part of our Universe, so $\dot R$ is the expansion
rate. During the pre-bounce epoch, $\dot r<0$, but if we choose this
to also be the dual phase of our bulk, i.e. $\tilde{R}=1/r$, we see
that $\dot{\tilde{R}}=-\dot r/r^2>0$.  If we take the post-bounce
epoch as corresponding to the normal phase $R=r$, then $\dot R=\dot
r>0$. Thus, by requiring that the T-duality phase transition happens
when the brane bounces off $r=r_0$, we have an Universe that always
expands.

Let us look at the two epochs in a bit more detail:
\begin{enumerate}
\item {\bf Pre-Bounce:} Since $\tilde R=1/r$, $\tilde H=-\dot r/r$
and therefore
\begin{eqnarray}
\dot{\tilde H}=-\frac{\ddot r}{r}+\frac{\dot
r^2}{r^2}&=&\frac{1}{2r}W^{\,\prime}(r)-\frac{1}{r^2}W(r).
\end{eqnarray}
In the region near the bounce, where $W\approx 0$ and
$W^{\,\prime}<0$, we have $\dot{\tilde H}<0$, typical of a power-law
inflationary scenario. The sign of $\dot{\tilde H}$ for large $r$
(i.e.\ very negative $\tau$), depends on the shape of the potential.
However, there is a very large class of potentials $W(r)$ such that
$\dot{\tilde H}$ can be made positive for large $r$. Such cases
imply a superinflationary Universe (i.e. accelerating with
increasing curvature) for large negative times. \item {\bf
Post-Bounce:} $R=r$ and $H=\dot r/r$, so $\dot H=-\dot{\tilde H}$.
$\dot H$ is therefore positive near $r=r_0$, and the curvature is
increasing. However, if $W(r)$ is such that $\dot{\tilde H}>0$ for
large $r$, then $\dot H<0$ in the same region, which in the
post-bounce scenario corresponds to large positive $\tau$. Hence, at
large times, the Universe is accelerating but its curvature is
decreasing.
\end{enumerate}
We see that, if the potential $W(r)$ has the right shape, we can
have a dual transition between a superinflationary Universe (the
pre-bounce epoch) and a post-inflationary, accelerating Universe
(the post-bounce epoch), characterized by the inversion of the scale
factor and a reflection of the equation of state. Remarkably, these
features are also typical of the Pre-Big Bang scenario
\cite{Gasperini}. Moreover, the transition between the two T-dual
phases occurs at a {\em finite} value of the scale factor,
corresponding to $r(0)=r_0$. This is reminiscent of some Pre-Big
Bang models, where the presence of a shifted dilaton potential in
the action avoids the formation of a singularity at the dual
transition \cite{Gasperini,Gasperini2,Gasperini3}. In particular, in
our model we can always tune the integration constants in order to
set $r_0$ equal to the self-dual radius, defined as the value of $r$
such that the scale factor and its dual are the same (in our
normalization units, this is simply $r_0=1$). It thus follows that
the Universe must have a minimum size determined by the self-dual
radius, which in turn is determined by the location of the zero of
the effective potential $W(r)$.

The transition between the dual Universes is characterised by an
interesting phase where the superinflation becomes power-law
inflation because of the change of sign of $\dot{\tilde H}$. Then,
at $\tau=0$, the dual transition occurs, the curvature begins to
increase, and the inflationary phase ends. Finally, the curvature
starts to decrease again and, at late times, the Universe eventually
enters our present epoch of accelerated expansion. Such a behaviour
depends entirely on the shape of the effective potential $W(r)$.

The effective potential discussed in Sec.~4 can be used as an
example for the scenario discussed above.  Indeed, the potential
(\ref{effpot}) vanishes at a point outside the event horizon, whose
location is invariant under T-duality transformations. Also, for
$r>r_0$, $W^{\,\prime}(r)<0$, hence the acceleration is always
positive.  However, it can be checked that $\dot H\rightarrow 0^+$
for $r\rightarrow\infty$, so the curvature increases at large times
towards a constant value, while the Universe is still accelerating.
Hence, the corresponding pre-Big Bang dual phase describes a
power-law inflationary model at all negative times. We believe that
this model can be improved by choosing a appropriate ansatz for the
dilaton field.  Indeed, the appearance of the effective potential
can be traced back to the junction conditions.
 In particular, Eq.~(\ref{juncbarphi}) can be written as
\begin{equation}
4\bar\phi^{\,\prime}\sqrt{1-B^2W}=\mu B,
\end{equation}
which reveals how crucial the choice for $\bar\phi$ is.

\section{Conclusions}\setcounter{equation}{0}
The results presented in this paper might build a bridge between
brane cosmology an Pre-Big bang scenario, and offer new lines of
investigations which, we believe, are worth studying.  First of all,
the tensor-scalar action introduced in Sec.~2 represents a new class
of backgrounds which can be much more general than the one
considered here.  For example, it would be interesting to study in
more details the black hole solutions discussed in Sec.~2, and
analyze their thermodynamical properties in light of the self-T
duality of the action.

In Sec.~3 we imposed the invariance under T-duality of the
junction conditions and we found that we recover standard brane
cosmological equations.  These are invariant provided that the
energy momentum tensor of the matter transforms in an appropriate
way.  The simplest case was considered in Sec.~4, where we showed
that we can still recover realistic cosmological equations.
 However, this result was achieved by assuming a specific form for
the brane Lagrangean.  Hence, it would be interesting to explore
more general cases; for example, by adding a dilatonic potential
on the brane and/or without assuming a specific form for the
functions $f(\phi)$ and $z(\phi)$ defined in Sec.~3.

Finally, in Sec.~5 we proposed a model where the dual phase of the
cosmological equations arises from the bouncing of the brane off
of a zero of the effective potential $W(r)$.  We showed that we
can recover most of the features of the Pre-Big Bang scenario, and
that the dual transition can occur at a non-singular point.  In
particular, we discussed the possibility of modelling a pre-Big
Bang superinflationary Universe evolving into a (dual) post-Big
Bang accelerating Universe, through an eventual power-law
inflationary phase.  More generally, the behaviour of the brane
cosmological equations are critically influenced by the structure
of the bulk (in particular if there are singularities and/or
horizons) and of the effective potential.  Therefore it is
important to study more general solutions to the bulk equations of
motion, not only because they are interesting {\em per se}, but
also because they might provide a more realistic brane
cosmological model.

\section*{Acknowledgements}\setcounter{equation}{0}
We wish to thank O.\ Corradini (Bologna U.) for very helpful
suggestions and stimulating discussions, and M.\ Gasperini (Bari U.
\& INFN) and G.\ Veneziano (CERN) for useful comments on the first
version.

\end{document}